\documentclass[a4paper,12pt]{article}
\usepackage{epsfig}\parskip 5pt plus 1pt
\usepackage{amsmath}
\usepackage{amssymb}
\usepackage{amsfonts}
\usepackage{fleqn}
\usepackage{mathrsfs}
\usepackage{cite}

\def\be{\begin{equation}}
\def\ee{\end{equation}}
\def\bea{\begin{eqnarray}}
\def\eea{\end{eqnarray}}

\begin{document}

\bibliographystyle{OurBibTeX}

\begin{titlepage}

 \vspace*{-15mm}
\begin{flushright}
\today\\
\end{flushright}
\vspace*{5mm}

\begin{center}
{ \bf \Large Exceptional Supersymmetric Standard Models with
  non-Abelian Discrete Family Symmetry}
\\[8mm]
R.~Howl\footnote{E-mail: \texttt{rhowl@soton.ac.uk}.} and
S.F.~King\footnote{E-mail: \texttt{sfk@hep.phys.soton.ac.uk}.}
\\
{\small\it
School of Physics and Astronomy, University of Southampton,\\
Southampton, SO17 1BJ, U.K.\\[2mm]
}
\end{center}
\vspace*{0.75cm}

\begin{abstract}
\noindent

We introduce a non-Abelian discrete $\Delta_{27}$
family symmetry into the recently proposed classes of
Exceptional Supersymmetric Standard Model ($E_6$SSM) based
on a broken $E_6$ Grand Unified Theory (GUT)
in order to solve the flavour problem in these models
and in particular to account for tri-bimaximal neutrino mixing.
We consider both the minimal version of the model
(the ME$_6$SSM) with gauge coupling unification at
the string scale and the E$_6$SSM broken via the
Pati-Salam chain with gauge coupling unification at
the conventional GUT scale. In both models there are
low energy exotic
colour triplets with couplings suppressed by the symmetries of the
model, including the family symmetry. This leads to suppressed
proton decay and long lived TeV mass colour triplet states
with striking signatures at the LHC.
\end{abstract}

\end{titlepage}
\newpage

\section{Introduction}
For more than thirty years the Standard Model has provided the most
accurate theoretical description of particle physics and, at present,
there is little direct experimental evidence to suggest that this
model should be replaced with a new theory.  But, despite being
experimentally sound, it is widely acknowledged that the model is
theoretically unsatisfactory in a number of areas \cite{Chung:2003fi}.
For instance, a large amount of fine-tuning is required to stabilize
the Higgs mass at the electroweak scale when the cut-off is taken to
be a high energy scale such as the Planck scale. There is also a lack
of explanation for the observed structure of the quark and lepton
masses and CKM matrix elements, no explanation for the observed small
neutrino masses and bi-large mixing angles, and, perhaps most importantly, the
Standard Model is incompatible with General Relativity, our most
accurate theory of gravity.

The most popular solution to the fine-tuning problem of the Higgs mass
is to treat the Standard Model as a low energy effective field theory
approximation to the Minimal Supersymmetric Standard Model (MSSM)
\cite{1}.  This can also potentially explain what dark matter consists
of and implies that there is unification of the Standard Model
forces within the framework of a Supersymmetric Grand Unified Theory
(SUSY GUT) at around $10^{16}$ GeV.
However, the MSSM does not entirely free the
Standard Model of problems with the Higgs mass since it introduces the
$\mu$-problem \cite{Martin:1997ns} (the unexplained origin
of the SUSY Higgs-Higgsino mass parameter with a TeV scale value)
and, because no superpartners have
been experimentally observed so far, a small fine-tuning problem.
Another related problem of SUSY GUTs is the question of how to split the
colour triplet Higgs apart from their Higgs doublet partners, giving
GUT scale masses to the former and weak scale masses to the latter,
while also satisfying the colour-triplet induced proton decay
experimental bounds.

An elegant solution to the $\mu$-problem is to extend the particle content
of the MSSM by introducing a Standard Model singlet
$S$ that couples to the Higgs doublets such that its dynamically
generated vacuum expectation value (VEV) provides an effective
TeV scale $\mu$-term that is related to the breaking of Supersymmetry
\cite{Ellis:1988er}. In such theories there is also some
advantage to be gained by having
an additional low energy Abelian gauge group factor $U(1)'$.
Without a $U(1)'$ gauge group a Goldstone boson
would be created by the singlet's VEV since the extended MSSM
superpotential has an associated global $U(1)$ symmetry
\cite{Kim:1983dt}. Alternative ways to resolve the would-be
Goldstone boson problem certainly exist, namely
the global $U(1)$ symmetry can be explicitly broken in
some way e.g. by adding an $S^3$ term to the superpotential,
as in the Next-to(N)MSSM \cite{Djouadi:2008uw}.
However, such approaches are always accompanied
by additional problems, for example, the $S^3$ term
introduces dangerous domain walls when a $Z_3$ discrete
symmetry associated with the NMSSM superpotential is broken.
By contrast the $U(1)'$ gauge group eats the Goldstone boson,
resulting in an observable massive $Z'$.

Standard Model (SM) singlets and $U(1)'$ gauge groups that can resolve
the $\mu$-problem of the MSSM as discussed above turn out to be
naturally contained within SUSY GUTs based on an $E_6$ gauge group.
In this paper we concentrate on $U(1)'$ subgroups of $E_6$ for which
the right-handed neutrinos are singlets so that a conventional see-saw
mechanism can be used. $E_6$ models that
contain a $U(1)'$ for which the right-handed neutrinos are singlets
have been collectively called Exceptional Supersymmetric Standard
Models.  Here we study two such models: the ME$_6$SSM (Minimal
Exceptional Supersymmetric Standard Model) \cite{Howl:2007zi} and the
usual E$_6$SSM \cite{King:2005jy}. Note that in both versions of the
model the TeV scale spectrum involves the matter content
of three complete $27$ supermultiplets of $E_6$
in order to cancel all the gauge anomalies family by family.
This means that, compared to the MSSM, there are an additional
three families of extra states with the quantum numbers of
three $5+\overline{5}$ representations of $SU(5)$ at the TeV scale.
These states will obviously ameliorate the little fine-tuning of the MSSM,
since they increase the lightest Higgs mass considerably \cite{King:2005jy}.

In the usual E$_6$SSM \cite{King:2005jy}
the $U(1)'$ gauge group, called $U(1)_N$, is a
combination of the $U(1)_{\psi}$ and $U(1)_{\chi}$ abelian subgroups
of $E_6$ defined by $E_6 \supset SO(10) \times U(1)_{\psi}$ and
$SO(10) \supset SU(5) \times U(1)_{\psi}$.  The combination of the
groups is chosen so that the right-handed neutrinos do not transform
under $U(1)_N$.  To cancel gauge anomalies for this group, three
copies of a $27$ supermultiplet of $E_6$ survive to low energies in
the model.  On top of this, two additional electroweak doublets (which have opposite $U(1)_N$ charges) are added at the TeV scale so that
unification of the Standard Model gauge coupling constants occurs at
the GUT scale.  Since the colour triplets are light in the E$_6$SSM, the proton
decay operators must either be forbidden or highly suppressed.
The former option is achieved using an exact $Z^B_2$
or $Z^L_2$ symmetry under which the colour triplets
are leptoquarks or diquarks \cite{King:2005jy}.
Such symmetries do not commute with the
$E_6$ (or its $SU(5)$ or Pati-Salam subgroups) and so the model is
written in terms of its Standard Model
representation.\footnote{Alternatively the theory can be written in
terms of several split $27$ multiplets so that the $Z^B_2$
or $Z^L_2$ symmetries do
commute with $E_6$ \cite{King:2005jy}.}

The ME$_6$SSM is a `minimal' version of the $E_6$SSM.  This refers to
the fact that the ME$_6$SSM does not contain the two additional
electroweak doublets required for unification of the gauge coupling
constants at the GUT scale, which can reintroduce a $\mu'$-problem
analogous to the original $\mu$-problem
\cite{Howl:2007zi}.  Gauge coupling unification is instead predicted
to occur close to the Planck scale in the ME$_6$SSM using an
intermediate Pati-Salam symmetry that is broken to the Standard Model
at the conventional GUT scale.  The $U(1)'$ group of the ME$_6$SSM,
denoted by $U(1)_X$, is not the same as the $U(1)_N$ but still
allows for a conventional see-saw mechanism since the right-handed
neutrinos remain neutral under it.
Unlike the E$_6$SSM, the proton decay
operators are highly suppressed rather than forbidden.
Since the $E_6$ symmetry is predicted to reside at the scale
at which quantum gravity effects should dominate,
the model is formulated in terms of the intermediate
Pati-Salam and $U(1)_{\psi}$ gauge groups.

In this paper we shall consider the E$_6$SSM as being broken
via the Pati-Salam chain as in the ME$_6$SSM. In this case
the only difference between the two models is that
the E$_6$SSM involves an additional
two low energy electroweak doublets, leading to
unification at the GUT scale.
Since, in the case of such an E$_6$SSM, the Pati-Salam gauge group
does not survive for very long before it is broken, the
phenomenology of the two alternative breaking chains
for the E$_6$SSM (Pati-Salam or $SU(5)$) is very similar,
differing only by the discussion of triplet decay and proton decay.
For the  E$_6$SSM as broken via the Pati-Salam chain,
the triplet decay and proton
decay discussion is the same as in the ME$_6$SSM.
When the $SU(2)_R$ and
$SU(4)_{PS}$ Pati-Salam gauge coupling constants are equal to each
other (as in the Pati-Salam version of the E$_6$SSM)
then the charge of the $U(1)_X$ group becomes
equivalent to the charge of the $U(1)_N$ group (see \cite{Howl:2007zi}
for a detailed explanation of this).

Despite their obvious attractions, as outlined above,
none of the E$_6$SSM models
so far proposed addresses the flavour problem,
i.e. provides an explanation for the structure of quark and
lepton masses and mixing angles.
In the past decade, the flavour problem has
been enriched by the discovery of neutrino mass and mixing,
leading to an explosion of interest in this area
\cite{King:2007nw}. A common approach is to suppose that the
quarks and leptons are described by some
family symmetry which is spontaneously broken at a high
energy scale \cite{gflsym}.
In particular, the approximately tri-bimaximal
nature of lepton mixing provides a renewed
motivation for idea the Yukawa couplings are
controlled by a spontaneously broken non-Abelian family symmetry
which spans all three families, for example $SU(3)$
\cite{deMedeirosVarzielas:2005ax,su3refs}, $SO(3)$ \cite{King:2005bj},
or one of their discrete subgroups \cite{deMedeirosVarzielas:2005qg,deMedeirosVarzielas:2006fc}.
In such models tri-bimaximal neutrino mixing arises from
a combination of vacuum alignment and (constrained) sequential
dominance \cite{King:1998jw}.
Furthermore, such family symmetries provide a solution to the
SUSY flavour and CP problems \cite{Ross:2002mr}.

The purpose of the present paper is to extend the above classes of
E$_6$SSM models to include a discrete non-Abelian family symmetry as a
step towards solving the flavour problem in these models. In
particular, we shall use the $\Delta_{27}$ family symmetry introduced
in \cite{deMedeirosVarzielas:2006fc} ($\Delta_{27}$ is a discrete non-Abelian subgroup of $SU(3)$). This is convenient since the
$\Delta_{27}$ family symmetry model in
\cite{deMedeirosVarzielas:2006fc} and the ME$_6$SSM in
\cite{Howl:2007zi} are both based on a high-energy Pati-Salam
symmetry. Following this approach we can also construct models based
on the E$_6$SSM with a $\Delta_{27}$ family symmetry which are broken
through the Pati-Salam chain, as
discussed in the previous paragraph.  The detailed strategy we shall
pursue is as follows.  We will introduce the $\Delta_{27}$ family
symmetry from \cite{deMedeirosVarzielas:2006fc} to the intermediate
Pati-Salam symmetry of the ME$_6$SSM or E$_6$SSM to build a model
based on a $\Delta_{27} \times G_{4221}$ gauge group where $G_{4221} \equiv \times SU(4)_{PS}
\times SU(2)_L \times SU(2)_R \times U(1)_{\psi}$.  The
resulting model can explain the observed mixing angles and mass
spectrum of the quarks and leptons, provide a tri-bimaximal mixing for
the neutrinos, solve the $\mu$-problem and small fine-tuning problem,
and does not involve doublet-triplet splitting.  A novel feature of
the ME$_6$SSM and the Pati-Salam formulation of the E$_6$SSM is that
proton decay is suppressed in a new way by the assumed $\Delta_{27}$
family symmetry and an $E_6$ singlet.  We also show how the
$\mu'$-problem can be solved in the E$_6$SSM using the $E_6$ singlet
that gets an intermediate VEV and suppresses proton decay.

The layout of the remainder of the paper is as follows.
In Section 2 we propose a model based on the
Pati-Salam gauge groups of the ME$_6$SSM or E$_6$SSM
and the
$\Delta_{27}$ family symmetry model. In Section 3 we
discuss gauge coupling unification at
the GUT scale in the E$_6$SSM or at the string scale in the
ME$_6$SSM.
Section 4 summarizes our results.

\section{ME$_6$SSM or E$_6$SSM with a $\Delta_{27}$ Family Symmetry}

In this section we introduce a $\Delta_{27}$
family symmetry into the ME$_6$SSM or E$_6$SSM
broken via the Pati-Salam chain.
The E$_6$SSM model with $\Delta_{27}$ family symmetry
has gauge coupling unification at the GUT
scale, rather than the string scale.
The resulting models are very powerful since they can
address the observed mixing angles and mass
spectrum of the quarks and leptons, including the tri-bimaximal mixing for the
neutrinos, the $\mu$-problem and the little fine-tuning problem of the
MSSM. We also show how the model solves the problem of
rapid proton decay (and colour triplet decay) without introducing
doublet-triplet splitting.

In the ME$_6$SSM or E$_6$SSM
the Standard Model quarks and leptons come from three
copies of the fundamental $E_6$ multiplet of dimension $27$.  Each
$27$ multiplet breaks into the following Pati-Salam representations:
$27 \rightarrow F + F^c + h + \mathcal{D} + S$ where $F,F^c$ contain
one generation of the leptons and quarks (and a charge conjugated
neutrino), $h$ can contain the MSSM Higgs bosons, $\mathcal{D}$ is
often called a colour triplet Higgs since it transforms as a colour triplet,
and $S$ is a singlet of the Standard Model.  The explicit Pati-Salam
representations of these states are listed in Table 1.  Following the
ME$_6$SSM and E$_6$SSM we take the third copy of the $27$ multiplets to contain the
MSSM Higgs bosons, which we denote by $h_3$.
In the $\Delta_{27}$ family symmetry model in
\cite{deMedeirosVarzielas:2006fc} the three generations of the leptons
and quarks $F$, $F^c$ transform as triplets under the
$\Delta_{27}$ group, and the MSSM Higgs bosons
$h_3$ transform as a singlet.
The rest of the ME$_6$SSM and E$_6$SSM states from
the three $27$ multiplets are not considered in the family symmetry
model.  In Sections 2.2 to 2.5 we explain the chosen $\Delta_{27}$
assignments for these ME$_6$SSM or E$_6$SSM states,
which are summarized by
Table~1. The only distinction between the ME$_6$SSM and E$_6$SSM
is that the latter involves an additional pair of electroweak
doublets $h',\overline{h}'$ in order to achieve unification
at the GUT scale.

We now briefly explain the approach to understanding Yukawa
hierarchies and neutrino tri-bimaximal (TB) mixing via broken
family symmetry, vacuum alignment and constrained sequential
dominance (CSD) (for more details see \cite{deMedeirosVarzielas:2005ax}, \cite{su3refs},
\cite{deMedeirosVarzielas:2006fc}, \cite{King:1998jw}).
The family symmetry
is broken by extra Higgs scalars
called flavons, often denoted by $\phi$ and $\overline{\phi}$.  The flavons
typically couple to the SM matter fermions via heavy messenger fields
giving rise (upon integrating out the messenger sector) to effective
Yukawa operators proportional to powers of the flavon fields
suppressed by powers of the messenger mass $M$.  The effective Yukawa
couplings are then expressed in terms of ratios of flavon vacuum
expectation values (VEVs) $\langle \overline{\phi} \rangle$ to these messenger
mass scales $M$, which defines a set of expansion parameters $\varepsilon \equiv \langle \overline{\phi} \rangle/M$.
If the neutrino masses are assumed to originate from the seesaw
mechanism,
the TB mixing pattern receives a natural explanation by
means of the so-called constrained sequential dominance
mechanism.  The basic idea is that only
one right-handed (RH) neutrino contributes dominantly to the
atmospheric neutrino mass and thus the atmospheric mixing angle
corresponds to a simple ratio of Yukawa couplings of just the dominant
RH neutrino. One of the subdominant RH neutrinos is then assumed to
govern the solar neutrino mass, in which case the solar mixing angle
corresponds to another simple ratio of Yukawa couplings associated to
this RH state.  The TB mixing pattern can then be implemented by means of
simple constraints on the Yukawa couplings. Since these emerge from
flavon VEVs, CSD is then achieved from a proper vacuum alignment of
flavons in the family space, for example $|\langle \overline{\phi}_3 \rangle|
\approx (0,0,1)$, $|\langle \overline{\phi}_{23} \rangle| \approx (0,1,1)$,
$|\langle \overline{\phi}_{123} \rangle| \approx (1,1,1)$, up to phases.


\begin{table}
\footnotesize
\begin{center}
 \begin{tabular}{ | c |c |c | c| c| c| c|}
 \hline
Field & $\Delta_{27} $ & $SU(4)_{PS} \times SU(2)_L \times SU(2)_R \times U(1)_{\psi}$ & $U(1)_R$ & $U(1)$ & $Z_2$ &$Z^H_2$
\\ \hline
$F$ & 3& $(4,2,1)_\frac{1}{2}$ & 1& 0& + & -
\\ \hline
$F^c$ & 3 & $(\overline{4},1,\overline{2})_\frac{1}{2}$& 1 &0& + &-
\\ \hline
$h_3$ ; $h_{1,2}$ &1 & $(1,2,2)_{-1}$ & 0& 0&+ & + ; -
\\ \hline
$\mathcal{D}_{1,2,3}$ & 1& $(6,1,1)_{-1}$ & 0 &0&+&-
\\ \hline
$S_3$ ; $S_{1,2}$ & 1& $(1,1,1)_2$ & 2 & 0 &+&+ ; -
\\ \hline
\hline
$16_H = \overline{H}_R,~H_L $ & 3 &$(\overline{4},1,\overline{2})_{\frac{1}{2}},~ (4,2,1)_{\frac{1}{2}}$ & 0 & 0&+&+
\\ \hline
$\overline{16}_H =  H_R,~ \overline{H}_L$ & $\overline{3}$ &$(4,1,2)_{-\frac{1}{2}},~ (\overline{4},\overline{2},1)_{-\frac{1}{2}}$ & 0 & 0 &+ & +
\\ \hline
$M$ & 1& $(1,1,1)_0$ & 2& 0& + & +
\\ \hline
$\Sigma$ & 1 & $(1,1,1)_0$ & 0& 5& - & -
\\ \hline
\hline
$H_{45}$ & 1& $(15,1,3)_0$ & 0 & 2 & +&+
\\ \hline
$\phi_{123}$ &3& $(1,1,1)_0$ & 0 & -1& +&+
\\ \hline
$\phi_{3}$ & 3 &$(1,1,1)_0$ & 0 & 3& +&+
\\ \hline
$\phi_{1}$ & 3 &$(1,1,1)_0$ & 0 & -4& -&+
\\ \hline
$\overline{\phi}_{3}$ & $\overline{3}$ & $(1,1,2 \times 2)_0$ &  0 & 0& -&+
\\ \hline
$\overline{\phi}_{23}$ & $\overline{3}$ & $(1,1,1)_0$ &   0 & -1& -&+
\\ \hline
$\overline{\phi}_{123}$ & $\overline{3}$ & $(1,1,1)_0$ &   0 & 1& -&+
\\ \hline
\hline
$h'$; $\overline{h}'$& 1 & $(1,2,1)_{x}$ , $(1,2,1)_{-x}$& 1 & -5 & + & +
\\ \hline
\end{tabular}
\end{center}
\caption{\footnotesize This Table lists all the particles (excluding the messengers) contained in the ME$_6$SSM and E$_6$SSM with a $\Delta_{27}$ family symmetry model where the $E_6$ symmetry is broken via the Pati-Salam chain.  The $\Delta_{27}$ and $G_{4221}$ representations are given for each particle, as well as the assignments for the additional constraining symmetries $U(1)_R \times U(1) \times Z_2 \times Z^H_2$.  The $F,F^c,h_3,h_{1,2},D_{1,2,3},S_{3}$ and $S_{1,2}$ particles are expected to come from three copies of a $27$ multiplet of a broken $E_6$ symmetry, the $16_H + \overline{16}_H$ are considered to be remnants of $27_H + \overline{27}_H$ $E_6$ states, the $H_{45}$ is expected to come from a $650$ multiplet of $E_6$ or as a composite of additional $27 + \overline{27}$ states, and, with the exception of $\overline{\phi}_3$, the flavons are singlets of $E_6$.  The three copies of the $27$ are the same as those in the ME$_6$SSM and E$_6$SSM. The flavons, $H_{45}$ and $H_R$ are the same as the equivalent particles in \cite{deMedeirosVarzielas:2006fc}, and the $\overline{H}_R$, $M$ and $\Sigma$ particles are similar to the equivalent states in the ME$_6$SSM. In the E$_6$SSM family symmetry model there are also two additional electroweak
doublets $h'$ and $\overline{h}'$ which cause the gauge coupling constants to unify at the GUT scale and have $U(1)_{\psi}$ charges of $\pm x$ where $x$ is some real number.  These particles are not in the ME$_6$SSM with a $\Delta_{27}$ family symmetry model.}
\end{table}

The model is defined in Table~1.
In addition to the Pati-Salam, $\Delta_{27}$ and $U(1)_{\psi}$
symmetries, extra discrete and abelian symmetries must also be applied
to constrain the model into a realistic theory.
The model that we formulate here is most simply constrained using the
combined symmetries $U(1)_R \times U(1) \times Z_2 \times Z_2^H$,
where $U(1)_R$ is an R-symmetry that contains the R-parity of the MSSM
as a subgroup. The $U(1) \times Z_2$ symmetries are adapted from
\cite{deMedeirosVarzielas:2006fc} and the $Z_2^H$ from
\cite{Howl:2007zi}.
In the ME$_6$SSM the $E_6$ symmetry is assumed to be broken to its
Pati-Salam and $U(1)_{\psi}$ groups near the string scale $M_S$.\footnote{Note that we expect the $E_6$ symmetry to be
broken at the String scale $M_S$ rather than the Planck scale since
extra states from the $\Delta_{27}$ model lower the
scale of unification of the ME$_6$SSM somewhat (see Section 3.2).}
This intermediate Pati-Salam with $U(1)_{\psi}$ is then expected to be
broken near the conventional GUT scale to the Standard Model with a
$U(1)'$ gauge group called $U(1)_X$.  In the $\Delta_{27}$ family
symmetry approach one expects the $SU(4)_{PS}$ and $SU(2)_R$
groups of the Pati-Salam symmetry to be broken by different mechanisms
rather than the same one as in the ME$_6$SSM and, in Section 3.2, we
show that we expect the $SU(4)_{PS}$ and $SU(2)_R$ groups to be broken
at two different scales in the ME$_6$SSM with $\Delta_{27}$ family symmetry model, with $SU(4)_{PS}$ broken at
the conventional GUT scale $M_{GUT}$ by $H_R$ VEVs,
and $SU(2)_R$ broken at the
compactification scale $M_C$, where we assume  $M_C>M_{GUT}$.  For the E$_6$SSM with $\Delta_{27}$ family symmetry model we show in Section 3.1 that the $SU(2)_R$ and $SU(4)_{PS}$ groups must both be broken at the GUT scale so that, in this case, $M_C = M_{GUT}$.

In the next subsection (2.1) we briefly explain how the $\Delta_{27}$
family symmetry from \cite{deMedeirosVarzielas:2006fc} when applied to
the ME$_6$SSM or E$_6$SSM (broken via the Pati-Salam group)
can explain the quark and lepton masses and mixing
angles using the Yukawa interactions generated by the symmetry.


\subsection{Yukawa Interactions}

In the ME$_6$SSM and E$_6$SSM models considered here
the $F$ and $F^c$ transform as $\Delta_{27}$ triplets and $h_3$
transforms as a singlet. This forbids the superpotential term $Y_{ij} F^i
F^{cj} h_3$, where $i,j = 1 \ldots 3$ and $Y_{ij}$ are theoretically
undetermined Yukawa coefficients. Instead higher order terms are
allowed that effectively generate the Standard Model Yukawa
interactions but with the desired Yukawa coefficients dynamically
generated to give the observed CKM matrix and quark and lepton masses.
This is achieved by introducing new particles to the theory that
couple to the fermions and quarks via their $\Delta_{27}$ components and
break the family symmetry to nothing. These new particles are called
flavons and are singlets of the Standard Model gauge group.  Six such
particles are required and their $G_{4221} \equiv SU(4)_{PS} \times
SU(2)_L \times SU(2)_R \times U(1)_{\psi}$ and $\Delta_{27}$
representations, as well as their $U(1)_R \times U(1) \times Z_2
\times Z^H_2$ charges, are given in Table 1.  These symmetry
assignments are simply borrowed from
\cite{deMedeirosVarzielas:2006fc}.

The leading Yukawa terms allowed by the symmetries are \cite{deMedeirosVarzielas:2006fc}:


\begin{equation} \label{eq:Yfirst}
\frac{1}{M^2_R} F^i F^{cj} h_3 \overline{\phi}_{3i} \overline{\phi}_{3j}
\end{equation}

\begin{equation} \label{eq:YH45}
\frac{1}{M^3_R} F^i F^{cj} h_3 H_{45} \overline{\phi}_{23i} \overline{\phi}_{23j}
\end{equation}


\begin{equation} \label{eq:Y3rd}
\frac{1}{M^2_R} F^i F^{cj} h_3 ( \overline{\phi}_{123i } \overline{\phi}_{23 j } + \overline{\phi}_{123j } \overline{\phi}_{23 i })
\end{equation}


\begin{equation} \label{eq:YH452}
\frac{1}{M^5_R} F^{i} F^{cj} h_3 H_{45} (\overline{\phi}_{3 i} \overline{\phi}_{123 j } + \overline{\phi}_{3 j} \overline{\phi}_{123 i}) (\overline{\phi}_{123k} \phi^k_{1})
\end{equation}


\begin{equation} \label{eq:Ylast}
\frac{1}{M^6_R} F^i F^{cj} h_3 \overline{\phi}_{123i} \overline{\phi}_{123j} (\overline{\phi}_{3k} \phi^k_{123}) (\overline{\phi}_{3l} \phi^l_{123})
\end{equation}

where the Latin indices refer to the $\Delta_{27}$ symmetry, and $M_R$ is the mass of right-handed
messengers, which is explained below. The $H_{45}$ in Eq.\ref{eq:YH45}
and Eq.\ref{eq:YH452} is an $\Delta_{27}$ singlet that transforms as
$(15,1,3)_0$ under the $G_{4221}$ symmetry.  This particle gets a VEV
in the hypercharge direction generating the Georgi-Jarlskog factor for
Eq.\ref{eq:YH45} \cite{Georgi:1979df,Ross:2002fb}.  This sets $m_\mu
\sim 3 m_s$ at the family symmetry breaking scale, which, after
radiative corrections from the Grand Unified scale in an MSSM inspired
GUT, agrees well with experimental data.  Since right-handed neutrinos
have zero hypercharge, the $H_{45}$ also suppresses the neutrino mass
matrix.  This is necessary for tri-bimaximal mixing to come from the
$\Delta_{27}$ family symmetry \cite{deMedeirosVarzielas:2005ax}.

The high order superpotential terms given by
Eq.\ref{eq:Yfirst}-\ref{eq:Ylast} are assumed to come from
renormalizable, high-energy interactions involving heavy vector-like
particles that transform in the same way as the quark and lepton
fields under the $G_{4221}$ symmetry.  Such particles, called
messengers, are integrated out of the high energy theory to generate
the above suppressed superpotential terms.  To distinguish the Yukawa
matrices for the up and down quarks we require that the $SU(2)_R$
messengers dominate over the $SU(2)_L$ messengers and, for the correct
up and down Yukawa matrices, we require that the up and down
right-handed messengers have mass $M_u$ and $M_d$ related by $M_u \sim
\frac{1}{3} M_d$ \cite{su3refs,deMedeirosVarzielas:2005ax}.  This can be achieved within the
framework of Wilson-line breaking of $SU(2)_R$ at some
compactification scale \cite{su3refs}.  We use $M_R$ to denote the
right-handed messenger scale, which could be $M_u$ or $M_d$ depending
on the interactions involved.

The flavons $\overline{\phi}_3 + \phi_3$, $\overline{\phi}_{23} + \phi_1$ and $\overline{\phi}_{123} + \phi_{123}$ get VEVs of order $\epsilon_3 M_d$, $\epsilon_d M_d$ and $\epsilon^2_d M_d$ respectively.  The $\Delta_{27}$ components that get VEVs are given by the flavons' subscripts.  Putting these VEVs into Eq.\ref{eq:Yfirst}-\ref{eq:Y3rd} generates the following leading order up and down quark Yukawa matrices \cite{deMedeirosVarzielas:2005ax}:

\[
Y_u \propto
\left( \begin{array} {ccc} 0 & \epsilon^2_u \epsilon_d & -\epsilon^2_u \epsilon_d \\
                           \epsilon^2_u \epsilon_d & -2\epsilon^2_u \frac{\epsilon_u}{\epsilon_d} & 2\epsilon^2_u \frac{\epsilon_u}{\epsilon_d} \\
                            -\epsilon^2_u \epsilon_d & 2\epsilon^2_u \frac{\epsilon_u}{\epsilon_d} & \epsilon^2_3
\end{array} \right)
~~~~~~~~~~~~~~~~
Y_d \propto
\left( \begin{array} {ccc} 0 & \epsilon^3_d & -\epsilon^3_d \\
                           \epsilon^3_d & \epsilon^2_d & -\epsilon^2_d \\
                           -\epsilon^3_d & -\epsilon^2_d & \epsilon^2_3
\end{array} \right)
\]

If $\epsilon_3 \sim 0.5 -1.0$,
$\epsilon_u \sim 0.05$, $\epsilon_d \sim 0.13$, then,
after radiative corrections from a high energy scale, the above
matrices (and corresponding lepton matrices) are able to generate
quark and lepton masses and CKM values that are in good agreement with
the observed values once the corrections from the higher order operators Eq.\ref{eq:YH452} and Eq.\ref{eq:Ylast} are included \cite{Ross:2007az}.

It should be noted that in both the ME$_6$SSM and E$_6$SSM with $\Delta_{27}$ family
symmetry models the renormalization group equations (RGEs) will be different
from those in the MSSM since there are three copies of a
supersymmetric $E_6$ $27$ multiplet below the conventional GUT scale (and two additional electroweak doublets in the E$_6$SSM model)
rather than just the MSSM particle spectrum.  This is illustrated by
Figure 1 for the ME$_6$SSM with $\Delta_{27}$ family
symmetry model. The Yukawa terms in the $\Delta_{27}$ model
\cite{deMedeirosVarzielas:2006fc} were assumed to be formulated at the
GUT scale and, after running the assumed MSSM from the GUT scale to
the electroweak scale, the results agree with the observed quark and
lepton mixing angles and masses.  In the ME$_6$SSM and E$_6$SSM with $\Delta_{27}$ family symmetry models the running
effects will clearly be different, but we do not expect
the main features of the low energy spectrum to be qualitatively
very different.

\subsection{Majorana Interactions}

In the ME$_6$SSM and E$_6$SSM models considered here
the Pati-Salam symmetry is broken to the Standard
Model by particles that transform as $(4,1,2)_{-\frac{1}{2}} +
(\overline{4},1,\overline{2})_{\frac{1}{2}}$ under $G_{4221}$.  The
$(4,1,2)_{-\frac{1}{2}}$ particle, denoted by $H_R$, once it develops
its GUT scale VEV, gives mass
to the right-handed neutrinos using Planck suppressed operators
$\frac{1}{M_p} \lambda_{ij} F^{ci} F^{cj} H_R H_R$.  This
non-renormalizable term, together with the Yukawa interaction involving
the neutrinos, can explain the small mass scale of the neutrinos but
not the observed hierarchical structure of neutrino masses and large
mixing angles without setting the couplings $\lambda_{ij}$ by hand.
In the $\Delta_{27}$ family symmetry model the particles that give
mass to the right-handed neutrinos also transform as $(4,1,2)$ under
the Pati-Salam gauge group but are taken to transform as anti-triplets
under $\Delta_{27}$.  With this $\Delta_{27}$ assignment, the particles can
dynamically generate the observed hierarchical structure of neutrino
masses and a tri-bimaximal mixing.  Following the $\Delta_{27}$ family
symmetry model, we therefore take the $H_R$ particle to transform as
an anti-triplet of $\Delta_{27}$.  The Majorana interactions are then
given by \cite{deMedeirosVarzielas:2006fc}:

\[\frac{1}{M_R} F^{ci} F^{cj} H_{Ri} H_{Rj}\]

\[\frac{1}{M^5_R} F^{ci} F^{cj}  \overline{\phi}_{23i} \overline{\phi}_{23j} H_{Rk} H_{Rl} \phi^k_{123} \phi^l_3\]

\[\frac{1}{M^5_R} F^{ci} F^{cj}  \overline{\phi}_{123i} \overline{\phi}_{123j} H_{Rk} H_{Rl} \phi^k_{123} \phi^l_{123}\]
Together with the neutrino Yukawa matrix generated by
Eq.\ref{eq:Yfirst}-\ref{eq:Ylast}, the above interactions produce a
$U_{PMNS}$ matrix with tri-bimaximal mixing and a hierarchical
structure of neutrino masses
in agreement with the observed values
\cite{Plentinger:2005kx}. How this happens is discussed in
\cite{deMedeirosVarzielas:2006fc} and references therein,
and the details of this are identical for the present model.

\subsection{The $\mu$-term and colour triplet Higgs Mass}

Taking $S_3$ to transform as a singlet under $\Delta_{27}$ allows the
superpotential term $S_3 h_3 h_3$.  This term is also allowed in the
ME$_6$SSM and E$_6$SSM.  If $S_3$ obtains a vacuum expectation value
at the TeV scale, $S_3 h_3 h_3$ will become an effective $\mu$-term of
the MSSM with the desired value of $\mu$ for electroweak symmetry
breaking.  The $S_3$ VEV is expected to depend on the breaking of SUSY
\cite{Martin:1997ns}, thus resolving the $\mu$-problem of
the MSSM.

In addition to solving the $\mu$-problem of the MSSM, this model
will also resolve the little fine-tuning problem of the MSSM.  This
is because there are extra particles below the conventional GUT scale
of $10^{16}$ GeV that are not contained in the
MSSM. These extra particles are from the three copies of the $27$
$E_6$ multiplet and form two copies of a $5 + \overline{5}$ of the
$SU(5)$ subgroup of $E_6$, and one colour triplet Higgs particle.  Due to
Renormalization Group effects, the extra states increase the value of
the Yukawa coupling constant for $S_3 h_3 h_3$ at low energies, and
hence increase the mass of the lightest
CP even Higgs boson \cite{King:2005jy}.

Since $S_3$ is assumed to get a VEV at the TeV scale, this suggests
that the
$\mathcal{D}_{1,2,3}$ particles from the three copies of the $27$
multiplet should transform as $\Delta_{27}$ singlets, so they may
all acquire TeV scale masses. If instead we assumed them to be
$\Delta_{27}$ triplets then at least one of their masses
would be expected to be lower than the electroweak
symmetry breaking scale, in violation of the direct experimental limits.
This is because we would expect the effective couplings
$S_3 \mathcal{D}_{1,2,3} \mathcal{D}_{1,2,3}$,
with $S_3$ obtaining a VEV at the TeV scale, to have a strongly
hierarchical mass structure, as in the case of ordinary quarks, with at
least the first generation, $D_1$, possibly having a mass lower the electroweak
breaking scale.  Instead, with $\mathcal{D}_{1,2,3}$ as $\Delta_{27}$
singlets, they will all obtain TeV scale masses from the (unsuppressed)
superpotential terms $S_3 \mathcal{D}_{1,2,3} \mathcal{D}_{1,2,3}$.
Similarly, we take the first two generations of $h$ from the fundamental
$27$ multiplets, which we denote by $h_{1,2}$, to transform as
$\Delta_{27}$ singlets so that they obtain TeV scale masses from the $S_3
h_{1,2} h_{1,2}$ superpotential terms.\footnote{Note that the first two
generations of $h$ and $\mathcal{D}$ can fit inside a $10_{-1}$
multiplet of $SO(10) \times U(1)_{\psi}$, but the third generations
cannot due to opposite $Z^H_2$ parity assignments.  Also note that the required TeV scale VEV of $S_3$ implies an effective $\mu$-term of similar magnitude, leading to a slight tuning required for electroweak symmetry breaking.}

\subsection{Proton Decay and colour triplet Higgs Decay}

Here we show how the family symmetry
can help to suppress proton decay arising from the light colour
triplet exchange.
The Pati-Salam $\mathcal{D}_{1,2,3}$ particles, which we shall refer
to as colour triplet
Higgs, decompose to $D_{1,2,3} \equiv (3,1)_{-\frac{1}{3}}$ and
$\overline{D}_{1,2,3} \equiv (\overline{3},1)_{\frac{1}{3}}$
multiplets of the Standard Model and will cause proton decay unless
the effective interactions $D_{1,2,3} Q Q + \overline{D}_{1,2,3}
u^c d^c$ or $ \overline{D}_{1,2,3} Q L + D_{1,2,3} \nu^c d^c +
D_{1,2,3} e^c u^c $, which are allowed by the $E_6$ superpotential
$27^3$, are heavily suppressed or forbidden
\cite{King:2005jy,Howl:2007zi}.  These operators are always present in
GUTs and SUSY GUTs, see Raby in \cite{Yao:2006px}. However, in the exact
$\Delta_{27}$ symmetry limit, operators of the form
$\mathcal{D}FF$ and $\mathcal{D}F^cF^c$ are forbidden,
since $F,F^c$ are family triplets while $\mathcal{D}$ are family singlets.


Once the $\Delta_{27}$ family symmetry is broken however,
proton decay operators will reappear suppressed by flavon and
other VEVs, and it becomes a quantitative question whether
these operators are sufficiently suppressed.
With the $Z^H_2$ and $\Delta_{27}$
symmetries chosen as in Table 1, the only way to generate these
proton-decay inducing terms is from higher order terms involving
flavons (to repair the $\Delta_{27}$ symmetry), and the $E_6$ singlet $\Sigma$ (to repair the $Z^H_2$ symmetry).  Taking $\Sigma$
to have $U(1) = + 5$ and $Z_2 = -1$, the smallest suppressed proton
decay terms are:\footnote{Replacing $D_{1,2,3}$ with $h_{1,2}$, and $F^i F^j$ by $F^i F^{cj}$, in
Eq.\ref{eq:1stPdecay}-\ref{eq:lastPdecay} gives the least suppressed
FCNCs that are induced by the `non-Higgses' $h_{1,2}$
\cite{King:2005jy}.}

\begin{equation} \label{eq:1stPdecay}
\frac{1}{M_S M^6_d} \Sigma \mathcal{D}_{1,2,3} F^{i} F^{j}   \overline{\phi}_{123i} \overline{\phi}_{23j} (\phi^k_{123} \overline{\phi}_{3k}) (\phi^l_{1} \overline{\phi}_{3l}) ~+~ (F^{i,j} \rightarrow F^{ci,j})
\end{equation}

\begin{equation} \label{eq:lastPdecay}
\frac{1}{M_S M^6_d} \Sigma \mathcal{D}_{1,2,3} (\epsilon_{ijk} F^{ci} \phi^j_{123} \phi^k_3) (\epsilon_{lmn} F^{cl} \phi^m_1 \phi^n_3)  (\phi^l_{1} \overline{\phi}_{123l}) ~+~ (F^{i,j} \rightarrow F^{ci,j})
\end{equation}

These operators are suppressed by the square of a string scale $M_S$,
which we take to be of order $10^{17.5}$ GeV.  We assume that this
type of suppression can be achieved due to the fact that the
messengers that couple the $\Sigma$ particle to the $F^c F^c
D_{1,2,3}$ superpotential term are different to the messengers that
couple the flavons and $H_R$ to the quarks and leptons in the Yukawa
and Majorana interactions of Sections 2.1 and 2.2.  We assume that
the former messengers reside at the unification scale which we take to
be a string scale $M_S \sim 10^{17.5}$ GeV, see Section 3.2 for further
discussion.  The effective terms $F^c F^c D_{1,2,3}$ are then
suppressed by a factor of about $\epsilon^6_d \epsilon^2_3
\frac{<\Sigma>}{M_S}$, which, for $\epsilon_d \sim 0.13$, $\epsilon_3
\sim 0.8$, $<\Sigma> \sim 10^{11}$ GeV, and $M_S \sim 10^{17.5}$ GeV,
is around $10^{-12}$.  This level of suppression should be just sufficient
to prevent proton decay from being observable in present experiments
if the colour triplets have mass greater than about $1.5$ TeV
\cite{Howl:2007zi}.\footnote{In \cite{Howl:2007zi} it was calculated
that the level of suppression required to prevent proton decay was
roughly $10^{-8}$ rather than $10^{-12}$.  The suppression of
$10^{-8}$ used in \cite{Howl:2007zi} only prevents proton decay if the
grand unified coupling constant for the interactions between the
colour triplets and the up and down quarks was of the same order of
magnitude as the up and down Yukawa coupling constant in the Higgs
sector.  This is not possible in the ME$_6$SSM with family symmetry
model however since the up and down Yukawa coupling constants are
generated by the flavon structure. We therefore require a suppression
of $\sim |Y_{u,d}| \times 10^{-8} \sim 10^{-12}$ for the appropriate
interactions.} We emphasize that the $10^{-12}$ level of suppression
is only a rough order of magnitude calculation and can be determined
from a number of sources, for example, the $d = 6$ proton decay
operators in R-parity violating models \cite{Barbier:2004ez}, and the
$d = 6$ proton decay operators in Grand Unified Theories with
doublet-triplet splitting \cite{Nath:2006ut}.  The present
experimental limit on the $d=6$ proton decay operator $p \rightarrow
\pi^0 e^+$ is $5.0 \times 10^{33}$ yrs \cite{Yao:2006px}.

To prevent the colour triplets from decaying with a lifetime smaller
than 0.1 s the interactions $F F \mathcal{D}_{1,2,3} + F^c F^c
\mathcal{D}_{1,2,3}$ should be suppressed by no more than roughly
$10^{-12}$ or $10^{-13}$ (using order of magnitude calculations from
\cite{Howl:2007zi}).  A lifetime longer than about 0.1s for the colour
triplets could cause problems for nucleosynthesis. The amount of Yukawa
suppression for these interactions is thus uniquely set to be about
$10^{-12}$ with the upper limit set
by the proton decay and the lower limit set by
colour triplet decay requirements.
This small allowed window of couplings warrants
a more detailed analysis of both proton decay and triplet
decay, which we hope will be performed in the future,
since it will lead to testable predictions for proton decay.
The long lived TeV scale colour
triplet states, which will be quasi-stable at colliders, lead to
striking signatures at the LHC \cite{Nisati:1997gb}.

The above solution to triplet-Higgs-induced proton decay is very
different from the solution used in conventional SUSY GUTs.  Generically
the solution is to make $\mathcal{D}_{1,2,3}$ very heavy (usually
above the GUT scale) using doublet-triplet splitting.  However, no
such doublet-triplet splitting is allowed in this theory since gauge
anomalies for the low energy $U(1)_X$ gauge group would be created
\cite{Howl:2007zi}, and instead the proton decay is suppressed by the
symmetries of the model (in particular the $\Delta_{27}$ family symmetry).

\subsection{R-parity and $H_R + \overline{H}_R$ Mass}

Not all the components of $H_R$ and $\overline{H}_R$ obtain mass by
absorbing the broken Pati-Salam gauge bosons when they acquire vacuum
expectation values in the right-handed neutrino direction.  To give
the rest of $H_R$ and $\overline{H}_R$ (and $H_L$ and $\overline{H}_L$
from the $SO(10)$ multiplets $16_H$ and $\overline{16}_H$) mass, we
have included a singlet $M$ in Table~1.
This singlet is assumed to get a GUT scale VEV, giving
mass to $16_H + \overline{16}_H$ from the superpotential term $M 16_H
\overline{16}_H$.  Since $M$ carries a $U(1)_R$ charge of $+2$, its
VEV breaks $U(1)_R$ to an R-parity.  This R-parity is the same as that
in the ME$_6$SSM, which is a generalization of R-parity in the MSSM.
This R-parity keeps the LSP stable, thus providing a dark matter
candidate.

\subsection{$h'$, $\overline{h}'$ Mass in the E$_6$SSM}

In the E$_6$SSM, to prevent the two additional
electroweak doublets $h'$ and
$\overline{h}'$ from introducing
gauge anomalies for the $U(1)_N$ gauge group, they are assumed have opposite $U(1)_N$ charges.  These particles effectively reintroduce a $\mu'$-problem
since there is no simple mechanism that explains why these particles
have low energy masses.  Here we give the particles mass by assuming that the $E_6$ singlet $\Sigma$ couples to the $h'$ and $\overline{h}'$ through the
non-renormalizable term $(1 / M_S) \Sigma \Sigma h' \overline{h}'$
and obtains a vacuum expectation value at $10^{11}$ GeV.  This gives $h'$ and $\overline{h}'$ the correct scale of mass for gauge coupling unification to occur at the GUT scale (see the third reference in \cite{King:2005jy}).

The way in which $h'$ and $\overline{h}'$ transform under
the Pati-Salam, $U(1)_{\psi}$, and
other symmetries is presented in Table 1,
which contains the total particle spectrum of the
E$_6$SSM (and ME$_6$SSM) with $\Delta_{27}$ family symmetry.\footnote{In Table 1 $h'$ and $\overline{h}'$ are chosen to transform as $(1,2,1)_{x}$ and $(1,2,1)_{-x}$ Pati-Salam representations respectively where $x$ is a real number.  Such multiplets cannot be derived from $E_6$ multiplets.}

\section{Gauge Coupling Unification}

\subsection{Unification and Symmetry Breaking in the E$_6$SSM}

In this subsection we briefly discuss the pattern of symmetry
breaking for the E$_6$SSM with a $\Delta_{27}$ family symmetry
model. Adding the extra electroweak
states $h'$ and $\overline{h}'$ at the TeV scale to the three
copies of a $27$ causes the Standard Model gauge coupling
constants to unify at the conventional GUT scale but with a higher
value than the MSSM prediction for the unification gauge coupling
constant (see the third reference of \cite{King:2005jy}).  This of
course requires that the $SU(2)_R$ and $SU(4)_{PS}$ subgroups of
the $E_6$ symmetry be broken at the same scale (the GUT scale).
However, as discussed further in the following Section (3.2),
we expect the $SU(2)_R$ and $SU(4)_{PS}$ groups to be
broken by separate mechanisms so that the $SU(2)_L$ messengers and
up and down $SU(2)_R$ messengers have different masses but the
quark and lepton components of the $SU(4)_{PS}$ messengers have
the same masses, as required for Section 2.1.  To achieve this we
assume that, at the GUT scale, the VEV of the $H_R +
\overline{H}_R$ multiplets breaks $SU(4)_{PS}$ to $SU(3)_c \times
U(1)_{B-L}$, and the $SU(2)_R$ is broken to $U(1)_{\tau^3_R}$ by a
Wilson-line \cite{su3refs}.\footnote{In addition to breaking the
$SU(4)_{PS}$ symmetry, the VEV of the $H_R + \overline{H}_R$
multiplets will also mix the $U(1)_{B_L}$, $U(1)_{\tau^3_R}$ and
$U(1)_{\psi}$ groups to create $U(1)_Y$ and the $U(1)_N$ group of
the E$_6$SSM.  The $U(1)_N$ group of the E$_6$SSM is generated,
rather than the $U(1)_X$ group of the ME$_6$SSM, because the gauge
coupling constants of the Pati-Salam (and $U(1)_{\psi}$)
symmetries are equal at the symmetry breaking scale
\cite{Howl:2007zi}.}
  This will give the up $SU(2)_R$
messengers masses smaller than the GUT scale.  Therefore, to
compensate for the effect on the running of the Standard Model
gauge coupling constants caused by the up $SU(2)_R$ messengers
(which would upset unification), we would require additional
messengers below the GUT scale that, together with the up
$SU(2)_R$ messengers, form a \emph{complete} $10$
multiplet of $SU(5)$.  The messengers below the GUT scale would
increase the MSSM prediction for the value of the unification
gauge coupling constant but keep the unification scale as the
conventional GUT scale.  Of course too many messengers, and too
small messenger masses, would cause the Standard Model gauge
coupling constants to blow up before they unify.  Here we simply
assume that the minimal number of messengers required to generate
the correct quark and lepton masses and mixing angles does not
prevent the unification of the Standard Model gauge coupling
constants at the GUT scale.

\subsection{Unification and Symmetry Breaking in the ME$_6$SSM}

In this subsection we discuss the pattern of symmetry breaking for the
ME$_6$SSM with a $\Delta_{27}$ family symmetry model and, using two
simple toy models, demonstrate that gauge coupling unification at the
string scale could be possible.  In the ME$_6$SSM the $E_6$ symmetry
is assumed to be broken at the Planck scale to a left-right symmetric
Pati-Salam gauge group $SU(4)_{PS} \times SU(2)_L \times SU(2)_R
\times D_{LR}$ (a maximal subgroup of $SO(10)$) and an abelian gauge
group $U(1)_{\psi}$.  The left-right symmetric gauge group is then
broken to the Standard Model gauge group with an additional abelian
gauge group $U(1)_X$, which is a combination of the charge of the
$U(1)_{\psi}$ group, the diagonal generator $\tau^3_R$ of the
$SU(2)_R$ group, and the diagonal generator associated with the
$U(1)_{B-L}$ subgroup of $SU(4)_{PS}$ defined by $SU(4)_{PS}
\rightarrow SU(3)_c \times U(1)_{B-L}$.  This breaking is achieved by
the ME$_6$SSM equivalent to the $H_R + \overline{H}_R$ particles from
 gaining VEVs in the right-handed neutrino directions.  At
the scale of this symmetry breaking the gauge couplings of the abelian
groups $U(1)_{B-L}$, $ U(1)_{\tau^3_R}$ and $U(1)_Y$ must satisfy the
following equation \cite{Howl:2007zi}:

\begin{equation} \label{eq:alphaY}
\frac{5}{\alpha_Y} = \frac{3}{\alpha_{\tau^3_R}} + \frac{2}{\alpha_{B-L}}
\end{equation}

For the ME$_6$SSM this is equivalent to \cite{Howl:2007zi}:

\begin{equation} \label{eq:alphaY2}
\frac{5}{\alpha_Y} =  \frac{3}{\alpha_{2R}} + \frac{2}{\alpha_{4PS}}
\end{equation}
Using $\alpha_{4PS} = \alpha_3$ and $\alpha_{2L} = \alpha_{2R}$ from
the left-right symmetry, the above equation can be written solely in
terms of Standard Model gauge coupling constants.  The scale of the
Pati-Salam symmetry is therefore determined by running the Standard
Model gauge couplings up until they satisfy this boundary equation.
With three copies of a $27$ multiplet at low energies this scale is
found to be $10^{16.4}$ GeV at the two-loop order \cite{Howl:2007zi}.

When we include the $\Delta_{27}$ family symmetry to the ME$_6$SSM,
the pattern of symmetry breaking is likely to change from the above
discussion. This is because the $SU(2)_R$ and $SU(4)_{PS}$ groups must
be broken by separate mechanisms so that the $SU(2)_L$ messengers and
up and down $SU(2)_R$ messengers have different masses but the quark
and lepton components of the $SU(4)_{PS}$ messengers have the same
masses, as required for Section 2.1.  To achieve this we assume that
the VEV of the $H_R + \overline{H}_R$ multiplets breaks $SU(4)_{PS}$
to $SU(3)_c \times U(1)_{B-L}$, and that $SU(2)_R$ is broken to
$U(1)_{\tau^3_R}$ by a Wilson-line at some compactification scale
\cite{su3refs}.\footnote{One could alternatively consider the VEV of $H_{45}$ to break $SU(4)_{PS}$ to $SU(3)_{c} \times U(1)_{B-L}$.  This depends on whether the VEV of $H_{45}$ is chosen to be at a greater or smaller energy scale than the $H_R + \overline{H}_R$ VEV.  In \cite{deMedeirosVarzielas:2005ax} and (the second reference in) \cite{su3refs}, for example, the $H_{45}$ VEV is taken to be of order $3 M_{d}$ and $3 \epsilon_d M_{d}$ respectively.}   To prevent the messengers from altering the
running of the gauge couplings of the ME$_6$SSM to a large degree, we
expect that $M_d$ should be of order or greater than the $SU(4)_{PS}$ breaking scale.  It then follows that, to generate $M_u \sim 3 M_d$, the compactification scale at which $SU(2)_R$ is broken to $U(1)_{\tau^3_R}$ should be around three times greater than $M_d$.



The pattern of symmetry breaking in this case is thus
expected to proceed as follows: the $SU(2)_R$ group is broken to
$U(1)_{\tau^3_R}$ at a compactification scale $M_C$, which, along with
the $SU(4)_{PS} \times SU(2)_L \times U(1)_{\psi}$ symmetry, is broken
at a lower scale to $SU(3)_c \times SU(2)_L \times U(1)_Y \times U(1)_X$
by the $H_R + \overline{H}_R$ particles. We also expect the left-right
discrete symmetry to be broken since the left-handed messengers are
heavier than and right-handed messengers. In realistic models of the
ME$_6$SSM with $\Delta_{27}$ family symmetry, we therefore do not
expect the scale of $G_{4221}$ symmetry breaking to be determined
uniquely using Eq.\ref{eq:alphaY} since there is no longer a symmetry
that sets $\alpha_{\tau^3_R}$ equal to $\alpha_{2L}$ at this scale.

The $H_R + \overline{H}_R$ particles also transform under the
$\Delta_{27}$ family symmetry and get VEVs in the third component so that they
break the $\Delta_{27}$ symmetry at the same scale as the $G_{4211}
\equiv SU(4)_{PS} \times SU(2)_L \times U(1)_{\tau^3_R} \times
U(1)_{\psi}$ symmetry. The remaining part of the family symmetry,
which is a subgroup of $\Delta_{27}$, will be
broken by the VEV of the $\phi_{23} + \overline{\phi}_{23}$ flavons at
the scale $\epsilon_d M_d$ where the right-handed messengers mass
$M_d$ should be above the $\Delta_{27}$ symmetry breaking scale
otherwise wavefunction insertions of the invariant operator $\phi_3
\phi^{\dag}_3 / M^2_R$ on a third family propagator can spoil the
perturbative expansion if $<\phi_3> ~>~ M_R$ \cite{su3refs}.

\begin{figure}
\includegraphics[angle=0, scale=1.012]{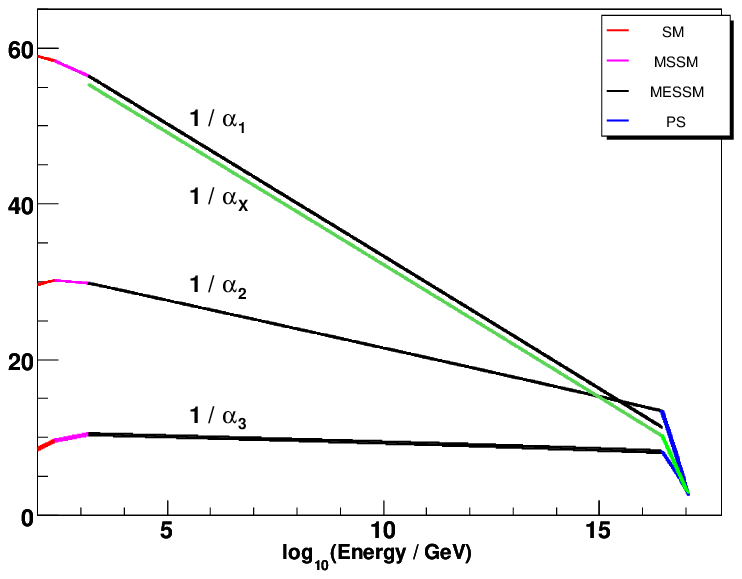} 
  \hspace{0.28 cm}
  \includegraphics[angle=0, scale=1.012]{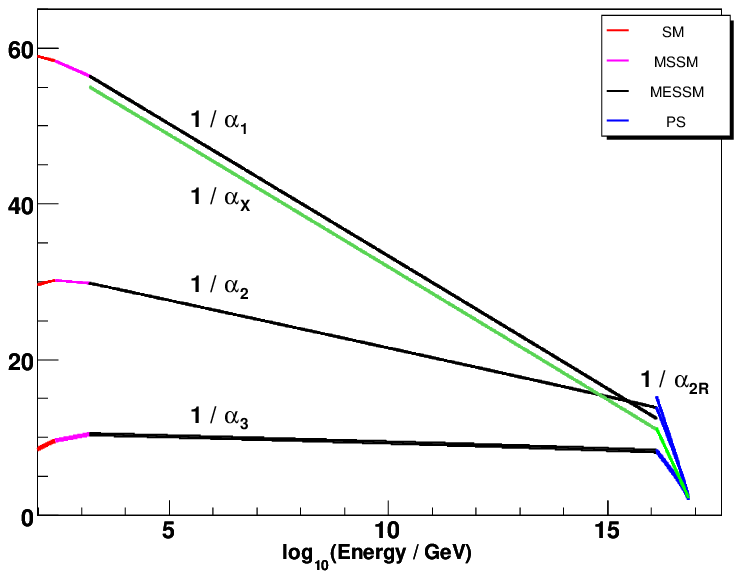} 
  \caption{\footnotesize
This figure illustrates the two-loop RGEs
  running of the gauge coupling constants for two models based on the
  ME$_6$SSM with $\Delta_{27}$ family
  symmetry.  It demonstrates that unification can be possible using a
  basic QFT description although we expect additional
  effects such as extra dimensions to change the running of the gauge
  couplings at higher energies and therefore to change the scale of
  unification.  The two models are described in detail in Section 3.2.
  The left panel is for a model with left-right
  symmetric intermediate $G_{4221}$ symmetry, whereas the right panel
  is for a more realistic non-symmetric $G_{4221}$ symmetry.  The
  scales of unification and $G_{4221}$ symmetry breaking are of order
  $10^{17.1}$, $10^{16.9}$ GeV and $10^{16.4}$, $10^{16.1}$ GeV for
  the left, right panel respectively.}  
   \label{b}
\end{figure}


The scale of the $E_6$ symmetry breaking in the ME$_6$SSM is also
expected to be modified when the $\Delta_{27}$ symmetry is included.
Instead of Planck scale $E_6$ symmetry breaking, we expect the $E_6$
symmetry to be broken at a string scale.  This is mainly due to the
number of additional particles (messengers) to the ME$_6$SSM states at
and above the $G_{4211}$ symmetry scale, which are required for the
$\Delta_{27}$ family symmetry to accurately describe the observed
quark and fermion masses and mixing angles.  These extra states cause
the gauge coupling constants to increase rapidly above the $G_{4211}$
symmetry breaking scale, bringing forward the unification scale.
Other modifications to the $E_6$ symmetry breaking scale in the
ME$_6$SSM will come from extra dimensions above the compactification
scale, the running of the gauge coupling constant for the abelian
$U(1)_{\tau^3_R}$ group, and the breaking of the left-right discrete
symmetry at the compactification scale.

Unification of the gauge coupling constants may in fact no longer be
possible when all of these changes from the ME$_6$SSM are calculated,
but in Figure 1 we demonstrate that gauge coupling unification still
occurs for two simple toy models of the ME$_6$SSM with $\Delta_{27}$
symmetry.
We make the approximation that
the compactification scale is equal to the $G_{4211}$ symmetry
breaking scale.  Both toy models therefore have an intermediate
$G_{4221}$ symmetry as in the ME$_6$SSM.  However, for the toy model
in the right panel of Figure 1, we assume that the left-right discrete
symmetry is broken at the unification scale due to the different
masses for the left-handed and right-handed messengers. Furthermore, we
also neglect any effects from extra dimensions above the
compactification scale.  In both panels of Figure 1 we assume that
three copies of an $E_6$ $27$ multiplet, which contain all the MSSM
states as well as new (non-MSSM) states, have mass at low energies
and, following the ME$_6$SSM, we take effective MSSM and non-MSSM
thresholds of $250$ GeV and $1.5$ TeV respectively.

At the $\Delta_{27} \times G_{4221}$ symmetry breaking scale, we also assume additional particles that break the symmetry and play a part in the $\Delta_{27}$ family symmetry's description of quark and lepton masses.  In the left panel we take these extra particles to consist of all the $G_{4221}$ states from five copies of $27 + \overline{27}$ multiplets, except for the $(6,1,1)_{\frac{1}{2}} + (6,1,1)_{-\frac{1}{2}}$ states which we assume have mass at the unification scale, as well as all the flavons given in Table 1 and a left-handed partner for $\overline{\phi}_3$.  The additional $27 + \overline{27}$ states contain the $16_H +\overline{16}_H$ particles that break the $\Delta \times G_{4221}$ symmetry and provide the Majorana interactions, the $16 + \overline{16}$ particles that give the $H_{45}$ as a composite, and messengers that also transform as a $16 + \overline{16}$ of $SO(10)$.  We assume that $H_{45}$ is a composite of a $16 + \overline{16}$ state since a fundamental $H_{45}$ particle (and its left-handed partner) would affect the running of the $SU(4)_{PS}$ gauge couplings by an amount that causes it to blow up before any unification of gauge couplings is possible, unless a large number of $SU(2)_L \times SU(2)_R$ extra states are added to compensate for this.  We would also need to explain why the rest of the $650$ $E_6$ multiplet, that contains the $H_{45}$, have larger mass.  On top of the five copies of the $27 + \overline{27}$ multiplets we also add additional Higgs messengers that transform as a triplet and an anti-triplet of the $\Delta_{27}$ family symmetry. These are required for unification of the gauge coupling constants.

For the right panel we include the same states as the left panel but without the left-handed  messengers as these are expected to get much larger masses than their right-handed components.  The scales of unification and $G_{4221}$ symmetry breaking are at $10^{17.1}$, $10^{16.9}$ GeV and $10^{16.4}$, $10^{16.1}$ GeV for the left, right panel respectively.  Note that the $G_{4221}$ symmetry breaking scales are close to the Grand Unification scale in conventional GUTs, we thus denote the scale by $M_{GUT}$.

We emphasize that these toy models do not represent accurate predictions for the running of the gauge coupling constants of the ME$_6$SSM with $\Delta_{27}$ family symmetry and are only used to demonstrate that, with the inclusion of the $\Delta_{27}$ messenger states to the ME$_6$SSM, gauge coupling unification is still possible but at a scale that is closer to the String scale than the Planck scale.





\section{Summary}

In this paper we have discussed models based on broken $E_6$ GUT with
a $\Delta_{27}$ (a discrete subgroup of $SU(3)$) family symmetry
broken close to the GUT scale.
To provide realistic models we also require additional
symmetries, including an R-symmetry which results in a conserved R-parity.
The models combine the ME$_6$SSM and E$_6$SSM proposed in
\cite{King:2005jy,Howl:2007zi} with the
$\Delta_{27}$ family symmetry approach of
\cite{deMedeirosVarzielas:2006fc}. The resulting synthesis
is very powerful and predictive, and solves a number
of problems facing the MSSM, including the
little fine-tuning problem, the $\mu$-problem and the flavour problem.
The solution to the $\mu$-problem requires an additional
low energy $U(1)_X$ gauge group, under which right-handed neutrinos
are neutral, allowing a conventional see-saw mechanism.
The $\Delta_{27}$ accounts for the quark and lepton masses
and mixing angles, with tri-bimaximal neutrino mixing
resulting from vacuum alignment and constrained sequential dominance.
Note that
we have considered both the ME$_6$SSM and E$_6$SSM formulated in terms of a
Pati-Salam symmetry (and an Abelian gauge group $U(1)_{\psi}$) yielding
the Standard Model gauge group (and an Abelian gauge group $U(1)_X$)
below the conventional GUT scale.
In the case of the E$_6$SSM the gauge group
$U(1)_X$ is identical to $U(1)_N$ of \cite{King:2005jy} since the
gauge couplings are unified at $M_{GUT}$.

The main phenomenological difference between the E$_6$SSM discussed here
(broken via the Pati-Salam chain) and the E$_6$SSM discussed in
\cite{King:2005jy} (broken via the $SU(5)$ chain)
arises from the physics of the colour triplet Higgs couplings.
In the original E$_6$SSM
\cite{King:2005jy},
exact $Z_2^L$ or $Z_2^B$ symmetries are allowed corresponding to
the colour triplet states coupling as diquarks or leptoquarks,
effectively preventing proton decay, while allowing
rapid colour triplet decay. However,
the Pati-Salam symmetry assumed here for both the ME$_6$SSM and E$_6$SSM,
prevents the use of the $Z_2^B$ or $Z_2^L$.
Instead, in both the ME$_6$SSM and E$_6$SSM,
the colour triplet Yukawa couplings must be suppressed
down to the level of $10^{-12}$, as required to sufficiently suppress
proton decay whilst allowing the states to decay before nucleosynthesis.
This is achieved by the symmetries of the model, with the $\Delta_{27}$
family symmetry playing
an important role in helping to achieve the required degree of suppression.
The highly suppressed couplings imply
long lived TeV mass colour triplets, with a lifetime typically about
0.1 sec for example, providing a striking
signature of these models at the LHC.


\subsection*{Acknowledgments}
RJH acknowledges a STFC Studentship.
SFK acknowledges partial support from the following grants:
PPARC Rolling Grant PPA/G/S/2003/00096;
EU Network MRTN-CT-2004-503369;
EU ILIAS RII3-CT-2004-506222;
NATO grant PST.CLG.980066.

\end{document}